Title: Peakspan: Defining, Quantifying and Extending the Boundaries of Peak Productive Lifespan

**Authors:** Alex Zhavoronkov, PhD [1,2,3 *]; Dominika Wilczok [4,5]

[1] Insilico Medicine AI Limited, Masdar, Abu Dhabi, UAE
[2] Insilico Medicine US Inc, Cambridge, MA 02138, USA
[3] Insilico Medicine Hong Kong Ltd., Hong Kong, Hong Kong SAR
[4] Duke University, Durham, NC27708, USA
[5] Duke Kunshan University, Jiangsu, Kunshan China

**Corresponding author**: Alex Zhavoronkov, PhD. alex@insilico.com

Abstract

The unprecedented extension of the human lifespan necessitates a parallel evolution in how we quantify the quality of aging and its socioeconomic impact. Traditional metrics focusing on Healthspan (years free of disease) overlook the gradual erosion of physiological capacity that occurs even in the absence of illness, leading to declines in productivity and eventual lack of capacity to work. To address this critical gap, we introduce Peakspan: the age interval during which an individual maintains at least 90% of their peak functional performance in a specific physiological or cognitive domain. Our multi-system analysis reveals a profound misalignment: most biological systems reach maximal capacity in early adulthood (20s–30s), resulting in a Peakspan that is remarkably short relative to the total lifespan. This dissociation means humans now spend the majority of their adult lives in a "healthy but declined" state, characterized by a significant "functional gap." We argue that extending Peakspan and developing strategies to restore function in post-peak individuals is the functional manifestation of rejuvenative biomedical progress and is essential for sustained economic growth in aging societies. Recognizing and tracking Peakspan, increasingly facilitated by artificial intelligence (AI) and foundational models of biological aging, is crucial for developing strategies to compress functional morbidity and maximize human potential across the life course.

Introduction

The doubling of human life expectancy over the past century represents a triumph of biomedical and societal progress. Yet, this demographic shift presents profound economic and societal challenges, particularly concerning productivity and the sustainability of retirement systems in rapidly aging populations. To understand the economic implications of longevity, it is useful to conceptualize three critical points in the human lifespan, as proposed by Zavoronkovs and Scerbakovs (2013): the conventional retirement age, the age of death, and the "Biological Retirement Age" (BRA), which is the point at which an individual can no longer perform adequately useful work due to functional decline [1]. The mismatch between these milestones, particularly the widening gap between BRA and the age of death, creates significant economic and social tensions. To delay the onset of BRA, the focus in geroscience and longevity biotechnology has appropriately shifted toward extending healthspan, defined as the duration of life free from debilitating chronic disease and disability [2]. However, healthspan is predicated on a morbidity-focused model, addressing the absence of disease rather than the presence of optimal function. It overlooks the gradual decline in functional capacity that characterizes normal aging, driven by the accumulation of molecular and cellular damage[3]. An individual can be clinically "healthy" yet operate significantly below their physiological or cognitive zenith. The historical focus of biomedical science has resulted in the extension of lifespan primarily by extending this "healthy but declined" period, rather than preserving youthful function [4–6]. This dissociation highlights a critical void in our ability to quantify the trajectory of human capability leading up to biological retirement.

To address this, we propose and formally define Peakspan: the period of the lifespan during which a person sustains at least 90% of their peak functional performance in a given domain. Peakspan begins when maximal attainable function is achieved and ends when age-related decline causes performance to fall below the 90% threshold. By age 50, a healthy individual has likely exited the Peakspan for most physiological functions, including fluid cognition, maximal aerobic capacity, pulmonary function, and thymic output, yet may still have 20 or more years of Healthspan remaining. This creates a substantial "Peakspan-Healthspan Gap" : the difference between current maximal performance and peak capacity. This gap manifests in subtle ways long before disease is diagnosed: slower reaction times, reduced endurance, or increased susceptibility to novel infections.  This functional gap is the primary target for restorative interventions. As we will argue, Peakspan is vital but overlooked in economic sustainability.

This publication defines and quantifies Peakspan across major physiological systems, explores its economic implications, and outlines the role of AI in extending it.

**Quantifying Peakspan**

The quantification of Peakspan relies on high-quality data regarding the trajectory of specific functions across the lifespan. Ideally, Peakspan should be determined using comprehensive longitudinal studies, which track the same individuals over decades. Such studies are essential for mitigating cohort effects (e.g., generational differences in nutrition, education, or environmental exposures) that can confound cross-sectional data. However, longitudinal data spanning the entire life course are rare and logistically challenging to collect. Consequently, Peakspan estimates rely on large-scale cross-sectional studies or meta-analyses of shorter longitudinal cohorts. When interpreting these data, it is crucial to acknowledge the vast individual variability. Genetics, sex, and lifestyle factors (nutrition, physical activity, socioeconomic status) significantly modulate both the absolute peak achieved and the rate of subsequent decline.

Table 1 summarizes the estimated Peakspans across core human functional domains, as well as key metrics and estimated scopes of decline based on the literature analysis for each selected system provided for each system below. The comparison of Peakspans and rates of decline for each system is visualized in figures 1.

Table 1. Estimated Peakspans (≥90% of Maximum Capacity) Across Major Human Organ Systems

| System | Peakspan | Time of Most Rapid Decline | Key Metrics |
|---|---|---|---|
| Cognitive | Fluid: 20-30; Crystallized: 45–54 | Fluid: 30–40 years (gradual acceleration); Crystallized: ~80 years | Processing speed; Working memory (digit span); Visual-spatial reasoning (block design score); Verbal comprehension/vocabulary (WAIS scaled scores) |
| Cardiorespiratory | 20s–25 | ≥40 years (gas-exchange acceleration) | $VO_2max$; Maximal cardiac output; HRmax; FEV1; FVC; DLCO |
| Musculoskeletal | 20s–35 | 60–70 years (torque/mass acceleration) | Grip strength; Muscle torque (e.g., knee extension); Bone mineral density; Lean muscle mass (DEXA/BIA) |
| Renal | 20s–30s | After ~45 years (GFR acceleration) | eGFR Creatinine clearance; Renal plasma flow; Kidney volume; Parenchymal thickness |
| Endocrine | 20s–30s | 30s–40s (inflections); Women: ~50 years (menopause); Men: Mid-adulthood onward (gradual) | Total testosterone (men); Estradiol (women); DHEA-S; Melatonin; FSH/LH |
| Sensory | 20s–30s | Mid-40s (vision); 20s (high-frequency hearing); 70s (somatosensation) | Pure-tone thresholds; Visual acuity; Contrast sensitivity; Olfactory identification score; Taste detection threshold; Vibration threshold |
| Immune | 20s-30s (Naïve T cells) | End of puberty for thymic exports | Naïve T-cell reserve, vaccine challenge performance, innate effector function |
| Digestive | 20–40s | ~40 years (liver, esophageal); Middle age (40s–50s, GI motility) | Hepatic clearance; Gastric emptying time; Colonic transit time; Fecal elastase; Portal vein flow; Esophageal motility |
| Reproductive | 20s-25 woman; 20-35 men | Women: After 25; Men: After 40 years | Fecundability ratio; Ovarian follicle count; Sperm motility; Total motile count |

**Cognitive Peakspan:**

Cognitive aging reflects the differing vulnerabilities of the brain regions to age-related processes. Abilities linked to fluid intelligence, including processing speed, visual–spatial reasoning, reasoning, and working memory, follow a consistent pattern of early optimization and gradual decline starting in ages between 20 and 30 [7, 8]. Processing speed, visual–spatial reasoning, fluid reasoning reach their apex between ages 20 and 24, with working memory peaking slightly later, at ages 25–29 [9–11]. In contrast, functions associated with crystallized intelligence, such as verbal comprehension and vocabulary, display a delayed but extended trajectory of growth and stability: these abilities peak between ages 45 and 54, remain relatively stable through the early seventies, and decline noticeably only around age 80 [7, 10, 12]. Consequently, although a single, unified cognitive Peakspan cannot be defined, cognition can be meaningfully divided into two distinct categories: a fluid intelligence Peakspan, characterized by an early apex and shorter duration of optimal performance, and a crystallized intelligence Peakspan, marked by a later peak and prolonged resilience against age-related decline.

**Cardiorespiratory Peakspan**:

Cardiovascular aging similarly reflects domain-specific vulnerabilities. A cardiorespiratory performance Peakspan, capturing maximal aerobic capacity ($VO_2peak/VO_2max$), maximal cardiac output, and HRmax, peaks in late adolescence to the mid-20s and then declines steadily at about 10% per decade [13, 14]. Maximal heart rate declines roughly linearly from early adulthood at ≈0.7 beats per minute per year constraining peak cardiac output despite training status [15]. On the respiratory side, lung function measured by Forced Expiratory Volume in 1 second (FEV1) and Forced Vital Capacity (FVC) peaks around 20–25 years [16]. Gas-exchange efficiency measured by DLCO shows age-related reductions that accelerate from ≥40 years in population follow-ups, approximately 0.025 mmoL/min/kPa/year [17, 18].

**Musculoskeletal Peakspan:**

Muscle strength peaks at 20-35 and plateaus at 35-50, with significant decline beginning at 65 years [19]. Peak bone mass density for women occurs in mid 30ies, and in 20-30 for men [20, 21]. Peak knee extension and flexion muscle torque was achieved at age 20 years across both sexes, with significant declines beginning as early as 20–30 years in men and 40–50 years in women, accelerating further between 60–70 years [22]. Grip strength for women plateaus until mid 50ies and declines steadily, while for men, the decline begins around mid 40ties [23]. Bilateral ankle dorsiflexion force-control deteriorates substantially with age, with older adults showing ~30–50% impairments across accuracy, variability, complexity, and coordination [24]

**Renal Peakspan:** Key renal health parameters like clearance of endogenous creatinine (Ccr) and estimated glomerular filtration rate (eGFR) is beginning to decline in early 30ies [25, 26]. Direct mGFR data show a nonlinear drop: ≈4 mL/min/1.73 m² per decade up to ~45 years, then ≈8 mL/min/1.73 m² per decade thereafter, consistent with longitudinal syntheses that place the normal decline for healthy adults between ~0.37 and 1.07 mL/min/1.73 m² per year [27]. Effective renal plasma and blood flow falls earlier, by roughly ~10% per decade beginning in the third–fourth decades, indicating hemodynamic reserve already wanes after ~30–40 years [28]. Structural metrics mirror these functions: kidney size begins to shrink after the fourth decade, with length falling ~20–30% and volume ~40% by the eighth decade [29] and 10% decrease in renal parenchymal thickness per decade [30].

**Endocrine Peakspan:**

Endocrine Peakspan is largely concentrated in early adulthood, with axis-specific inflection points emerging by the 30s–40s. The hypothalamic–pituitary–gonadal axis shows sex-specific timing: in men, total testosterone decreases ~0.8–1% per year (free/bioavailable ~2%/yr) beginning in mid-adulthood, so a man at 60 typically has ~20–30% lower total testosterone than at 40 [31]. In women, estradiol remains cyclic through the 30s–40s but declines during perimenopause and plummets across the menopausal transition (median menopause ~50–51 years), reaching low postmenopausal levels thereafter [32]. Adrenal androgens peak by the late teens/20s, plateau until mid 30ies, plummet until 40ies and then decline steadily [33]. Serum IGF-1 in the late 70s is roughly 30–35% of the value observed in young adulthood, and continues to decline [34].

**Sensory Peakspan:**

Sensory Peakspan is early–mid adulthood, with domain-specific inflection points emerging by the 30s–40s. Hearing is most sensitive in the 20s; extended high-frequency thresholds (>8 kHz) already worsen in the 20s [35]. Regarding vision, although greatly variable at an individual level, it seems to start declining mid-40s as contrast sensitivity, accommodation and acuity exhibit progressive deficits [36]. Olfaction peaks at 20–30 years and then gradually decreases [37]. Gustation declines more modestly and heterogeneously: detection thresholds for basic tastes in older adults are on average 2-5 times higher than in younger adults [38]. Somatosensation peaks in the 20-30ies, but the decline becomes statistically significant only in the 70ies [39]

**Digestive Peakspan:** Digestive Peakspan appears to favor early–mid adulthood with organ-specific inflection points. In the liver, hepatic unbound clearance falls ~0.8% per year beginning at ~40 years (≈−32% by age 80 vs 40), consistent with age-related reductions in liver weight and blood flow; portal-vein 4D-flow MRI further suggests a hemodynamic peak around ~45 years, with ~40% lower net portal flow by ages 60–69 versus younger adults [40]. Gastrointestinal motility exhibits earlier, region-specific shifts; gastric and small-intestinal transit were already faster by middle age, while colonic transit was significantly slower in middle-aged women (but not men), indicating sex-dependent large-bowel slowing that emerges before 50 [41]. Esophageal motor performance also shows age-sensitive degradation detectable from the fifth decade as HRM series and clinical cohorts report higher odds of motility abnormalities beyond ~40 years [42]. By contrast, exocrine pancreatic output is comparatively preserved through midlife in the general population, with clinically low fecal elastase uncommon before 60 but present in roughly 10–22% of community-dwelling elders ≥70–80 years, underscoring that substantial pancreatic exocrine insufficiency typically appears later than hepatic or esophagogastric changes [43].

**Immune Peakspan**:

Thymic involution after puberty causes naïve T-cell output to collapse rapidly, with thymic export falling to about 20% of its pre-puberty level by age 25 and to ≈ 5 % by the age of 55 [44]. Naïve T cells and T memory stem cells peak in early adulthood (18–29 years) before steadily declining with age, while central and effector memory T cells progressively increase to a peak plateau at age 64 years before stabilizing beyond 65 years [45]. The B-cell Peakspan extends from late adolescence to early adulthood; beyond 30 years, intrinsic molecular capacity for Ig diversification and memory cell formation progressively decreases, halving by midlife and declining to roughly one-quarter by old age [46].

**Reproductive Peakspan:**

Reproductive Peakspan is concentrated in the 20s–early 30s, with sex-specific inflection points evident by the mid-30s. In women, per-cycle fecundability is already measurably lower after 35: in a prospective cohort, fecundability at 35 was ~14% lower than at 30, with further declines at older ages and post-menopausal cessation [47]. Ovarian reserve follows a long, quantifiable decay both quantitatively and quality-wise; 30 year old women are estimated to have 12% of pre-birth non-growing follicles, while 40 year old women have 3% [48]. Consistent with this biology, meiotic errors and oocyte aneuploidy rise sharply after the mid-30s [49]. Men's reproductive aging is more gradual but detectable by ~35–40: peak semen quality evidenced by highest values in volume, sperm count, motility, total motility, vitality, occurred in the 18–39 years age span (average 33.1 years), with a sharp decline evident after age 40 years[50].

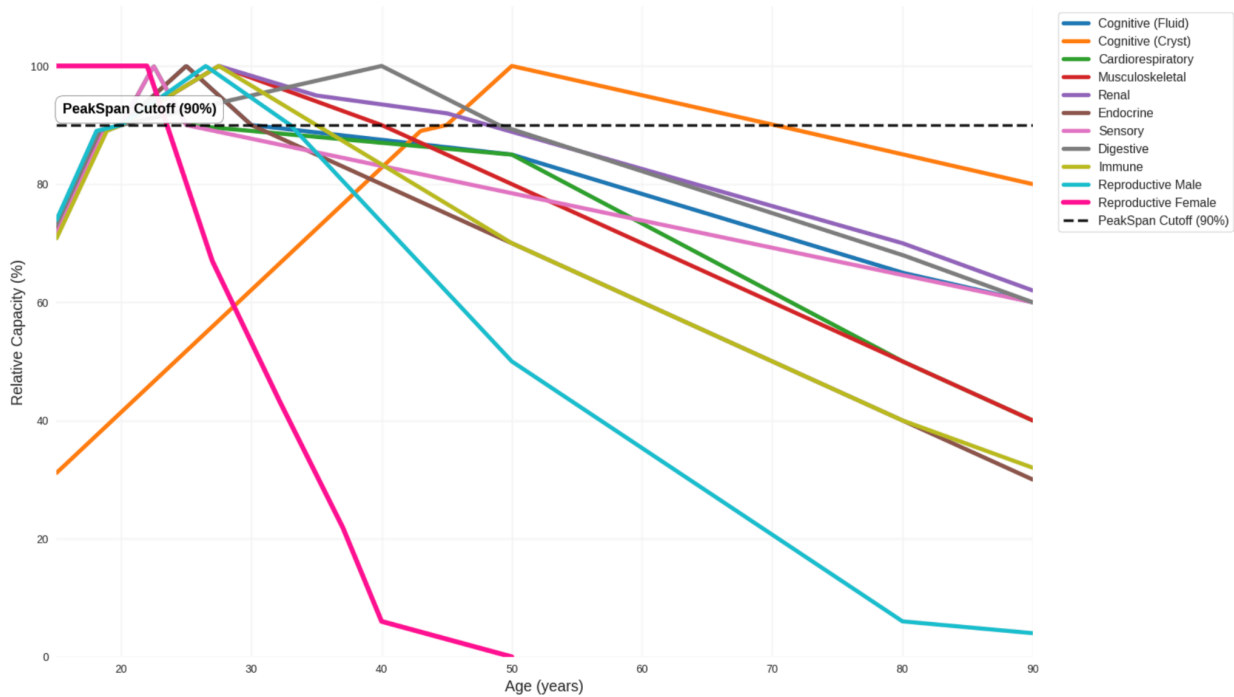

*Figure 1: Culminative graph of Peakspan across systems*

Figure 1 Legend: Comprehensive written analysis of peak and rate of decline of each selected system has been performed. For the graphic purposes, cardiorespiratory system decline was proxied by Vo2 max. The endocrine system was proxied by IGF1. Female reproductive capacity was proxied by the ability to conceive. Male reproductive system was proxied by total progressive motile sperm count. The renal system was proxied by eGFR. The sensory system was proxied by olfactory system performance. The digestive system was proxied by hepatic unbound clearance. Fluid and crystallized intelligence were proxied by WAIS scores. The musculoskeletal system was proxied by ankle dorsiflexion force-control.

**Artificial Intelligence for Peakspan Research**

Tracking Peakspan enables a proactive optimization of human function towards maximal capacity throughout the life course, with significant implications for clinical practice, research, and socioeconomic policy. AI integrates diverse data streams (clinical biomarkers, wearables, genomics), allowing for the creation of personalized functional trajectories [51]. Biological age quantification was accelerated and improved through "Deep Aging Clocks" [52–55]. Deep neural networks analyze high-dimensional data including blood biochemistry [54], transcriptomics [56], epigenetics [57, 58], imaging [59–61] or the microbiome taxonomy [62] to predict age accurately and indicate signs of accelerated/decelerated aging [63]. Modern aging clocks primarily model residuals from chronological age, assigning "younger-than" or "older-than" labels that do not specify system-level functional capacity. We propose a Peakspan-oriented reframing in which clocks estimate functional proximity to an individual's own peak performance rather than deviation from calendar age. In this framework, Peakspan clocks classify whether a given organ or system is operating within its person-specific peak window and quantify both the distance from that window and the rate of departure. Central to this reframing is the delta-peak age, a per-person alignment marker that registers functional readouts to the individual's true peak rather than to chronological age. Peakspan clocks thereby provide performance-anchored readouts suitable for trial endpoint selection, and monitoring in geroscience trials, emphasizing preservation of high-function person-time instead of marginal compression of late-life risk. Peakspan is tractable to model because most individuals reach peak function during lifetime and early-adult measurements are common across longitudinal cohorts, wearable data streams, athletic evaluations, occupational screening, and routine clinical testing. For Peakspan, the decision-relevant region is the functional edge; modeling this band is tractable because most individuals reach peak function and early-adult measurements are common in cohorts, wearable records, athletic assessments, occupational screening, and routine clinical testing. Peakspan therefore calls for models that recover system-specific maxima, estimate time spent near those maxima, and forecast the timing of slope transitions. A practical formulation is edge-focused learning and includes restricting training to observations within roughly 10% of each person's estimated peak and learn both short-horizon persistence within the near-peak band and the hazard of exiting it, while treating Peakspan as a joint problem of state estimation and change-point prediction. The base network maps molecular, physiological, imaging, and behavioral inputs could be normalized to functional state for each system, referenced to the individual's estimated peak rather than a population mean; a second output predicts time to the next inflection, supervised by change-point–aware fits to longitudinal trajectories in validated endpoints such as VO2 max, FEV1, DLCO, grip strength, eGFR, and contrast sensitivity. Calibration draws on age-stratified cohorts to anchor early-adult maxima and late-life plateaus; assay-aware normalization, batch correction, and explicit domain-shift penalties support transportability across studies and laboratories.

*Life models for Peakspan prediction*

The advent of transformer architectures has enabled the creation of foundational models trained on vast biological datasets to model the aging process itself. Such biological foundation model, dubbed "life model", learns joint embeddings across methylomes, proteomes, metabolomes, wearable time series, imaging-derived phenotypes, and routine laboratories after harmonization; training emphasizes cross-modal reconstruction and temporal consistency over repeated measures to prevent shortcut learning [64]. An interventional forecaster conditions on exposures relevant to peak preservation, including training volume, nutritional patterning, sleep regularity, vascular medications, and endocrine therapies, and produces short-horizon forecasts of organ-specific function with calibrated uncertainty. Monotonicity constraints encode physiological irreversibilities, for example non-increase in ovarian function after menopause absent exogenous replacement [65]. A cohort-level aggregator rolls individual trajectories into projections that support trial planning and RCT emulation [66]. External validity is tested by recovering established regularities like sex differences in $VO_2$max decline and by demonstrating stable calibration across decades. Reporting focuses on metric-level calibration and coverage, not only discrimination, so that small slope changes are credibly detected over two to three years. The insights generated by these AI tools can underpin personalized longevity guidance systems and could be further purposefully redesigned to preserve Peakspan, where small, actionable changes are most likely to prevent system-wide decline.

**Economic models for studying the effects of Peakspan extension**

A practical approach to integrating Peakspan into macroeconomic analysis is to place it in an overlapping-generations framework with human-capital accumulation and an age-specific efficiency profile. In this setting, aggregate output depends on quality-adjusted labor: hours supplied weighted by proximity to functional peak, calibrated to empirical measures of cognition, cardiorespiratory fitness, strength, and sensory performance. The sustainability of economic growth depends critically on maintaining the productivity of the workforce as the population ages. The economic impact of biomedical progress can be understood through the framework of Rejuvenative (RBMTR) versus Non-rejuvenative (NBMTR) progress rates [1]. NBMTR encompasses advances that extend lifespan or manage disease without restoring youthful function. Excessive reliance on NBMTR can increase the duration of dependency. RBMTR encompasses advances that actively restore or maintain youthful function effectively extending or restoring Peakspan. Sustained economic growth in aging societies depends critically on RBMTR outpacing NBMTR. If RBMTR > NBMTR, the productive capacity of the population increases, the BRA is delayed, and the economic burden of non-productive populations is reduced. If NBMTR dominates, lifespan extends without a commensurate extension of productive years, leading to economic stagnation or decline.

We propose that focusing societal investment specifically on extending Peakspan is the critical driver for significant economic growth. This perspective yields a simple investment rule. Prioritize interventions that demonstrably slow midlife erosion in fluid cognitive performance or cardiorespiratory fitness, or that postpone abrupt losses in performance-limiting domains. Report effects in economically interpretable units such as percentage-point changes in annual decline or years of postponement so that they can be imputed into age–productivity schedules and aggregated across cohorts to forecast output and welfare. Because the required inputs are slopes and timing, both directly estimable from longitudinal data, the translation from biomarker gains to macro impact is transparent and reproducible. The asynchrony of human capital strengthens the case as extending earlier-peaking capacities preserves adaptation and learning in midlife, while the durability of later-peaking expertise supports error-sensitive, experience-intensive tasks. Targeted Peakspan extension therefore reallocates time toward high-productivity states and aligns health investment with sustained economic performance rather than expanded dependency. Policies that extend Peakspan shift the distribution of worker-years toward high efficiency, raising effective labor input without increasing headcount or hours. The policy problem reduces to two parameters: the midlife slope of efficiency decay in the limiting system, and the timing of any rapid-decline threshold. Estimates from trials or longitudinal cohorts, expressed as percentage changes in annual decay rates or as years of postponement for breakpoints, feed directly into these parameters and generate revised age–efficiency distributions. A flatter midlife slope or a later breakpoint expands the near-peak share across cohorts, lifts lifetime earnings, reduces costly errors in precision-dependent work, and narrows the gap between conventional retirement age and functional capacity. These shifts delay the biological retirement age and compress the old-age dependency ratio by replacing low-efficiency years with high-efficiency years in the life-cycle labor profile.

**Discussion**

Peakspan organizes geroscience around the aim of keeping the system's functional capability at at least 90% of its maximum capacity for as long as possible. Cross-system evidence shows asynchronous trajectories. Fluid cognition and high-intensity aerobic capacity decline early. Female reproductive capacity is truncated by a discrete endocrine transition and is one of the few systems that go down to 0% functionality over time. Renal filtration and pulmonary gas exchange bend in midlife with acceleration later on. Crystallized abilities retain high plateaus. This pattern argues for timing prevention and treatment to system-specific inflections rather than applying a uniform schedule.

Peakspan should become a primary object of study in geroscience. The largest preventive leverage sits at the edge where function first drops below 90% of individual peak. Intervening at this boundary delays the cascade of cross-system decline that ultimately produces disability and dependence. An edge-first program specifies system-specific thresholds, monitors short-interval slopes, and tests interventions for their ability to maintain the near-peak band. This approach complements disease-centric endpoints by targeting the earliest functional losses that propagate into clinical morbidity and economic dependency. Therefore, in geroscience research, Peakspan provides quantitative, functionally relevant endpoints that can be tracked on shorter timescales than disease incidence or mortality. An effective intervention should not only slow decline but ideally restore function towards the peak.

The economic framing clarifies clinical priorities. Gains in near-peak years carry outsized value in occupations that depend on rapid processing, reliable power output, or fine sensory discrimination. This creates feedback from economic benefit to biomedical development that is stronger than traditional cost-offset logic.

Important constraints should be explicit. Measurements are heterogeneous across cohorts and often non-linear in relation to perceived function, which complicates normalization to a common peak. Sex differences extend beyond reproduction and require sex-specific models and targets. Social determinants shift slopes well before biological inflection, which argues for causal adjustment and sensitivity analysis rather than post-hoc subgroup claims. External validation across cohorts and laboratories is mandatory, as is transparent uncertainty quantification when models are used for individual decisions or policy planning.

**Conclusion**

Peakspan is short for functions that govern capability and economic value. Progress in longevity biotechnology should be measured in added years above a defined functional threshold within the organs that limit performance. Orienting measurement and intervention toward Peakspan, not only healthspan, aligns geroscience targets with clinical action and macroeconomic benefit.

Moving forward, several avenues are critical for advancing the study and application of Peakspan:

1. **Integrating Peakspan into Research and Clinical Practice:** Longitudinal cohort studies must incorporate comprehensive functional assessments starting early in life to accurately map Peakspan trajectories. Clinical trials of anti-aging interventions must include functional endpoints to directly assess Peakspan extension and restoration. Healthy longevity medicine physicians should gather data to monitor the rates of decline in each system, and use geroscience tools to limit that decline.
2. **Leveraging Advanced AI:** The integration of AI, deep aging clocks, and wearable technology opens unprecedented possibilities for real-time, personalized tracking of Peakspan. The development of sophisticated foundational AI models, exemplified by the PreciousGPT framework [64–66] is crucial for modeling the aging process, simulating interventions, and discovering novel therapies.
3. **Advancing Restorative Therapies:** Significant investment is needed to translate promising restorative therapies from the lab to the clinic, with a rigorous focus on safety and efficacy.

Studying and extending Peakspan at the near-peak edge will yield the greatest returns for prevention, clinical practice, and economic sustainability. AI systems should be purpose-built for this task by training and evaluating models on the 90–100% peakspan for each system, where small slope changes are detectable on practical timescales and where most existing data already concentrate. Prioritizing this range converts incremental functional maintenance into longer productive years and reduced dependency, aligning biomedical progress with measurable gains in capability.

Peakspan provides a valuable lens to assess how well we are preserving human potential. It challenges us to aim higher than the mere absence of disease. By striving to align Peakspan more closely with Lifespan, we can ensure that longer lives are not just longer, but better filled with vitality, productivity, and capability up to the very end.

**Authors' contributions:** AZ - original idea and writing. DW -ideas, writing and editing.

**Conflicts of Interests:** AZ is affiliated with Insilico Medicine, a commercial company developing and using generative artificial intelligence and other next-generation AI technologies and robotics for drug discovery, drug development, and aging research. Insilico Medicine has developed a portfolio of multiple therapeutic programs targeting fibrotic diseases, cancer, immunological diseases, and age-related diseases, utilizing its generative AI platform and a range of deep aging clocks and AI life models. DW declares no conflicts of interests.